\documentclass[sw]{agutex}

\authorrunninghead{SCHRIJVER, C.J.}

\titlerunninghead{SPACE WEATHER: MILD, MODERATE, EXTREME }

\authoraddr{Corresponding author: C.J. Schrijver, Lockheed Martin Advanced Technology Center, 
Dept. A021S/B252, 3251 Hanover Street, Palo Alto, CA94304, USA. (schrijver@lmsal.com)}

\def\rr#1{#1}
\def\rr2#1{#1}

\begin{document}

\title{Socio-economic hazards and impacts of space weather: the important range between mild and extreme}

\author{Carolus J. Schrijver \altaffilmark{1}} \altaffiltext{1}{Lockheed Martin Advanced Technology Center, Palo Alto, California, USA}

\begin{abstract}
  Society needs to prepare for more severe space weather than it
    has experienced in the modern technological era. To enable that,
    we must both quantify extreme-event characteristics and analyze
    impacts of lesser events that are frequent yet severe enough to be
    informative.  Exploratory studies suggest \rr2{that economic impacts of a
    century-level space hurricane and of a century of lesser
    space-weather ``gales'' may turn out} to be of the same order
    of magnitude. The economic benefits of effective mitigation of the
    impacts of space gales may substantially exceed the required
    investments, even as these investments provide valuable
    information to prepare for the worst possible storms.
\end{abstract}

\begin{article}

\section{Societal concern about space weather}
Interconnected electronic and electrical technologies are rapidly
spreading their reach into our personal lives and into the overall
functioning of our global society. Among the backbones of this
ensemble are the electric power grid and the armada of Earth-orbiting
satellites. The global use of electrical power grows substantially
faster than the world's population: in the 15 years prior to 2012, the
world population grew by 20\%, while electric power use grew by 60\%\
\citep{iea2013}. The satellite industry grows even faster than that,
with an increase in global revenue by 420\%\ over the same period
\citep{sia2013}. As electronics become more powerful and miniaturized,
society discovers innovative new ways of making use of them, accessing
information through the internet and finding its way with a
satellite-based navigation system, while drawing electricity that is
generated in a multitude of power plants that are increasingly
interconnected by more extended and more complex grids of high-voltage
lines (e.g., \citet{jason2011}).

A century ago, space weather was hardly on anyone's radar, other than
as presenting scientific puzzles related to geomagnetic variability
and aurorae, and curious phenomena that sometimes occurred in
telegraph systems (e.g., \citet{2010hssr.book...15O}). Half a century
ago, space-weather perturbations of communications by radio waves or via submarine telephone cables were a substantial
but transient nuisance. Over the past quarter century, however, the
growth in our space-based assets and in the reliance on stable
electric power has made us very much aware of space weather as
something that is not limited to academic interest or of relevance to the potential future human exploration of the solar system, but that reaches
into our everyday lives, along with weather, earthquakes, and other
natural hazards. And, as for these other phenomena, the impacts can be
mild and noticed primarily in, among others, electric power quality or system
operations (e.g.,
\citet{forbesstcyr2008,forbesstcyr2010,2012SpWea..10.5001F}), may lead
to component failures by immediate or cumulative damage (e.g., \citet{gaunt2013}), or may be extensive
with the potential of being catastrophic (e.g.,
\citet{swximpactlloyds2011}).

This realization has led to an increased focus on the research of the
broad variety of Sun-Earth connections, to the creation of space
weather forecasting organizations around the world, and to a rapidly
growing awareness of space weather in our society. For example, at the
time of this writing, asking Google for web sites with the phrase
``space weather'' yields close to 2.8 million results. There is also a
rapid growth in the number of organizations and individuals interested
in receiving space weather alerts and warnings by email: the US NOAA
Space Weather Prediction Center alone currently serves over 42,000
subscribers with their ``product subscription service'' (PSS).

The growing awareness of the impacts of space weather on virtually any
sector in society appears founded in the realization that it is a real
and often-present hazard. A survey of the PSS subscribers showed that
about half had a professional (rather than personal) interest, and
that these represented a variety of government and private-sector
organizations of a wide range of scales
\citep{schrijverrabanal2013}. Some 60\%\ of these subscribers checked
current conditions and near-term forecasts at least once a day, often
expressing concerns about human safety and reliability or continuity
of services. Almost half of the respondents to the survey considered
the strongest class of space weather to have a very strong to strong
impact on their organizational operations, and about three quarters
stated that it was likely that their organization would somehow
respond to the forecast of an extreme space weather event, with many
expressing the expectation that this would mitigate the impact on the
operations.

\section{Extreme space weather and its impacts}
The US National Research Council (NRC) captured perspectives from a
broad spectrum of stakeholders in society in a report entitled
``Severe space weather events -- Understanding societal and economic
impacts'' \citep{severeswx2008}. The report notes that ``an
estimate of \$\,1 trillion to \$\,2 trillion during the first year alone
was given for the societal and economic costs of a 'severe geomagnetic
storm scenario' with recovery times of 4 to 10 years'', potentially
affecting more than 130 million people.

These hypothetical consequences for the U.S.\ were later quantified
for the global economy in an assessment by \citet{schulte2014}. They
analyzed the effects of three different scenarios for the main
impacted regions --~essentially broad local-time zones along
longitudinal axes centered on the Americas, on Europe and Africa, and
on East Asia and Australia~-- and then assessed the impacts on
different countries around the world to deduce an overall cost. They
assumed that a geomagnetic disturbance (GMD) that might occur once in
a century would cause the loss of 10\%\ of the power-generating
capacity in the impacted region for up to a year. With those
assumptions, they essentially confirmed the NRC report's hypothesis,
while quantifying the even larger global consequences:``the possible
impact of a century-scale space-weather event on the global economy
\ldots\ could be'' USD\,2.4 trillion ``up to USD 3.4 trillion or
5.6\%\ of global GDP, and impacts would affect sectors and populations
well outside the direct area of impact.'' They ``find that a severe
space-weather event could lead to global economic damage of the same
order as wars, extreme financial crisis, and estimated future climate
change.'' In other words, impacts of extreme space storms need to be
paid serious attention.

\section{The continuum of space weather impacts}
Without having experienced the anticipated technological consequences
of uncommonly severe space weather on modern-day technologies, the
impact estimates are based on assumptions about both the event and its
consequences. Assuming, with \citet{schulte2014}, that 10\%\ of the
power-generating capacity is lost for a year is obviously a
hypothetical impact. As to the magnitude of the space storm the NRC
report offers a rather vague definition not unlike what we see
elsewhere in the literature: ``Extreme space weather events are
low-frequency/high-consequence (LF/HC) events and as such present
--~in terms of their potential broader, collateral impacts~-- a unique
set of problems for public (and private) institutions and governance,
different from the problems raised by conventional, expected, and
frequently experienced events'' \citep{severeswx2008}. That definition
quantifies neither the magnitude, nor the occurrence frequency of
``extreme space weather events''.

When it comes to working through the physics of the impact chains and
the numerical assessment of impacts, a pragmatic choice
needs to be made: the characteristic ``century-scale space-weather
event'' is commonly equated to the flares and geomagnetic storm days
on September $1-2$ of 1859. These events were independently observed,
and reported on, by Richard Carrington and Richard Hodgson in England
\citep{1859MNRAS..20...13C,1859MNRAS..20...15H}, and remain even in
modern-day perspective among the strongest solar events directly
observed
\citep{2004SoPh..224..407C,2006AdSpR..38..119C,2013SpWea..11..585B}. But
are they really the extremes in the mathematical sense, i.e., could
nothing more intense happen?

A compilation of evidence of solar, stellar, terrestrial, and
lunar ``records'' (either direct observations or analyses of residual
proxies, such as chemical and radionuclide signatures in rocks, ice
deposits, and the biosphere) led \citet{2014EOSTr..95Q.201S} to conclude
that very much more energetic events could
occur on the Sun, with flare energies up to several hundred times
larger, and with integrated fluxes of particle storms at least a few times
larger, than the maximum values directly observed over the past
decades. With occurrence frequencies dropping with increasing event
strength, one needs to consider how the potential of such dramatic
events once per millennium or even once per million years should
factor into our preparation and mitigation activities.

What is immediately obvious from the available material on space
weather event intensities, however, is that there is no separation
between mild and extreme, but instead a continuum from the frequent
small events to the rare most violent explosions. It would be a
mistake to think that all we need to worry about is the
``century-level space weather event''. Rather, events with frequencies
measured in months to millennia all need to be considered.

The continuous spectrum of magnitudes of space weather storms is not a
surprise. We see the signature ``power laws'' in many aspects of the
chain of processes reaching from Sun to Earth, including in sizes of
solar active regions (e.g., \citet{harvey+zwaan93}); in solar flare
energies (as summarized by \citet{2014EOSTr..95Q.201S}), possibly
extended by statistics of very large flares observed in some stars
that appear to be like the Sun in terms of mass and rotation rate
(e.g., \citet{2014PASJ...66L...4N}); in the size (opening angle)
distribution of solar eruptions forming coronal mass ejections
(\citet{robbrecht+etal2009}, and references therein); in peak proton
intensities during solar energetic particle events (e.g.,
\citet{1975SoPh...41..189V}); in (a proxy for) power input into the
Earth's magnetosphere in solar-wind bursts
(\citet{2011GeoRL..3814111M}, and references therein); in the ``sizes,
energies, and durations of substorms'' (\citet{2010JGRA..11512217V},
and references therein); and in auroral emissions
(\citet{2010JGRA..115.6202K}, and references therein).

\section{Breezes, gales,  and hurricanes}
Why is it important to explicitly include the continuum of storm
strengths in our thinking along with the concerns about the extreme?  The analogy with winds in terrestrial
weather can help make a case. Breezes are pleasant and have generally
no detrimental impact and perhaps their benign effects outweigh the
adverse. Hurricanes are damaging, but occur so infrequently in any
given location that direct, personal experience is -~fortunately~-
limited.  It is from the intermediate gales that many can learn efficiently. With that
analogy in mind, I argue that:

(1) The impacts of fairly frequently experienced severe space weather events, when put in purely economic terms, appear not to be negligible when compared to those of a ``century-level
space-weather event''. In fact, allowing for the substantial uncertainties resulting from assumptions being made to arrive at the overall impact estimates, they appear to be comparable. For example, the
study by \citet{2014SpWea..12..487S} that assessed the cost of the
non-catastrophic impact of non-extreme geomagnetic disturbances on the
US alone through insurance claims attributable to the variations in
the electric power grid suggests an average economic impact (in 2015
equivalent units after correction for inflation) of order
USD\,$7-10$\,billion per year. {In the absence of evidence to the contrary, that estimate was reached assuming that perturbations induced by space weather are not substantially different from all others; that assumption is in obvious need of substantiation through detailed system studies that were beyond the data available to \citet{2014SpWea..12..487S}.}

Industry studies suggest that impacts of
electric-power quality and continuity variations are comparable for
the United States and for the European Union
\citep{primen2001,mansontargosz2008}. {If perturbations induced by space weather have similar impacts within the U.S.\ and E.U., then} 
the
overall impact through power-grid effects of ``common'' space weather,
not including possible impacts in Asia, integrated over a century
comes in at a total of USD\,$1.3 - 2.1$\,trillion. {Although this is a rough approximation that needs follow up, it appears}
worth society's while to pay attention to the full range of space
storms. We shall have to figure out how to parcel up that attention
between organizations that deal with disasters and those that deal
with day-to-day issues within our infrastructure, but we should not
neglect one aspect while dealing with the other.

(2) The occurrence and impact of events that happen a few times a year
to once in a decade offer valuable lessons, both in terms of
preparation for severe space weather and to learn about the workings
of the distant universe where what we call space weather envelopes
exoplanets that orbit a large fraction of the stars in the sky. For
example, work by \citet{2014SpWea..12..487S} (and references therein)
suggests that variability in electric power quality and its impacts on
electrical and electronic equipment increases by $\approx$20\%\ on the
5\%\ most geomagnetically active days. On the one hand, we can develop
and hone our scientific understanding, forecasting tools, and
mitigation protocols based on the relatively frequent sub-extreme
events. On the other hand, we can use such events to trace how the
variety of physical phenomena associated with space weather find their
way into our technological infrastructure. That knowledge can be
shared to create design criteria and best practices on how to make our
infrastructure less susceptible and more resilient. By investing in
both science and engineering studies, society will come out stronger
and better prepared to face the consequences of all manner of space
weather impacts. And if we do this well enough, it {may pay for
itself if the above estimates stand up to scrutiny}: avoiding even a fraction of the estimated impacts of
``common'' space weather could pay for the research and protective
infrastructures needed to be ready for ``extreme space weather
events''.

\section{In conclusion}
{In my opinion, the state of our knowledge of space weather and its
  socio-economic impacts is such that operational monitoring and
  scientific investments need to be advanced jointly. There is an
  urgent need to improve forecasts in order to reduce demonstrated
  impacts caused by severe space weather and the considerable concerns
  regarding more severe events. Yet key links between Sun and society
  are inadequately explored or understood. This leaves the end-to-end
  chain too weak for reliable, actionable, customer-specific
  forecasts. We need to look into the possibilities of the extreme
  events and into the impacts that occur frequently enough for both
  statistical and case studies.}

{\rr2{The ICSU Committee on Space Weather (COSPAR) and the steering committee of the International Living with a Star (ILWS) program appointed an international panel with the charge to develop} a roadmap that
outlines where the largest advances are needed on the research side,
and how they can be effectively made {by investing in infrastructure, partnerships, and agency collaboration} \citep{cosparrm2015}: 
\begin{itemize}
\item advance the Sun-Earth system observatory along with
  observation-guided models;
\item understand the early phases of space weather as this originates
  at the Sun and propagates into the heliosphere. \rr2{In order to enable this, the roadmap identifies suggestions} for new instrumentation to obtain
  binocular 3D coronal imaging, to measure magnetic energy storage and
  release throughout the lower solar atmosphere, and to provide
  additional perspectives well off the Sun-Earth line, all aimed at
  improving our ability to model the erupted solar magnetic field
  heading towards Earth;
\item understand the factors that control how geospace is primed for,
  and driven to, the generation of geomagnetically-induced currents
  (GICs) and harsh particle radiation as solar wind, magnetosphere,
  and ionosphere couple. \rr2{For this, the roadmap offers ideas} for new instrumentation to
  expand in-situ coverage of the auroral particle acceleration region
  and the dipole-tail field transition region that aid in the
  determination of the evolving magnetospheric state, and for
  instrumentation to complement satellite data of the magnetospheric
  and ionospheric variability to cover gaps; and
\item develop a comprehensive space environment specification
  throughout the Sun-Earth system by utilizing terrestrial, geospace,
  heliospheric, and astrophysical (from lunar to stellar) information.
\end{itemize}}

{The COSPAR/ILWS roadmap recommends that, in parallel to the
  above,} the global society needs to invest in forecasting and
awareness, and also in education of the general public and of
subject-matter experts in society's infrastructure, space-weather
operations, emergency management, and scientific research \rr2{(see \citet{cosparrm2015} for details and references, and also, for example, \citet{2014JSWSC...4A..05V}, and \citet{2014SpWea..12..530M})}. {This
  requires starting with the quantification of the vulnerability of
  humans and of society's infrastructure for space weather by
  partnering with user groups; building test beds in which coordinated
  observing supports model development; standardizing (meta-) data and
  product metrics, and harmonizing access to data and model archives;
  and realizing observational coverage of the Sun-society system.}

{Society is increasingly dependent on reliable electrical power,
  communication, and navigation systems. Paramount to sucessful
  preparation for severe and extreme space weather is that the
  strategies for the research and operational investments be developed
  in a broad and open discussion. This requires dialogues between
  stakeholders in research, operations, and society. And it requires
  careful consideration of those scientific advances that should most
  effectively advance overall readiness for space weather of all
  magnitudes, and of how scientific and operational platforms can best
  complement each other.}


\end{article}
\end{document}